\documentclass[conference]{IEEEtran}
\IEEEoverridecommandlockouts
\usepackage{cite}
\usepackage{amsmath,amssymb,amsfonts}
\usepackage{algorithmic}
\usepackage{graphicx}
\usepackage{caption}
\usepackage{textcomp}
\usepackage{xcolor}
\usepackage{epsfig}
\usepackage{subcaption}
\usepackage{calc}
\usepackage{graphicx}
\usepackage{wrapfig}

\usepackage{amsmath}
\usepackage{subcaption}
\usepackage{hyperref}

\usepackage{booktabs}
\usepackage{comment}
\usepackage{caption} 
\usepackage{times}
\usepackage{epsfig}
\usepackage{adjustbox}
\usepackage{multirow}
\usepackage{times}
\usepackage{colortbl}
\usepackage{epsfig}
\usepackage{graphicx}
\usepackage{adjustbox}
\usepackage{amsmath}
\usepackage{amssymb}
\usepackage[ruled,vlined,linesnumbered]{algorithm2e}
\usepackage{algorithm2e}

\def\BibTeX{{\rm B\kern-.05em{\sc i\kern-.025em b}\kern-.08em
    T\kern-.1667em\lower.7ex\hbox{E}\kern-.125emX}}
\begin{document}
\captionsetup[figure]{labelformat={default},labelsep=period,name={Fig.}}
\captionsetup[table]{
  labelsep=newline,
  justification=centering,
  singlelinecheck=false,
}
\title{Decentralized Collaborative Pricing and Shunting for Multiple EV Charging Stations Based on Multi-Agent Reinforcement Learning\\

}

\author{\IEEEauthorblockN{1\textsuperscript{st} Tianhao Bu}
\IEEEauthorblockA{ 
\textit{Glory Engineering \& Tech Co., LTD}\\
\textit{Shanghai, China} \\
tianhao.bu19@alumni.imperial.ac.uk}
\and
\IEEEauthorblockN{2\textsuperscript{nd} Hang Li$^{\star}$}

\IEEEauthorblockA{\textit{School of Electronic Information}  \\\textit{and Electrical Engineering} \\
\textit{Shanghai Jiao Tong University}\\
\textit{Shanghai,China} \\
lihang9596@sjtu.edu.cn\\$^{\star}$Corresponding author }
\and
\IEEEauthorblockN{3\textsuperscript{rd} Guojie Li}
\IEEEauthorblockA{\textit{School of Electronic Information}  \\\textit{and Electrical Engineering} \\
\textit{Shanghai Jiao Tong University}\\
\textit{Shanghai,China} \\
liguojie@sjtu.edu.cn}
\

}

\maketitle

\newcommand{\head}[1]{{\smallskip\noindent\textbf{#1}}}
\newcommand{\alert}[1]{{{#1}}}
\newcommand{\mich}[1]{\textcolor{blue}{#1}}

\newcommand{\sm}{\scriptsize}
\newcommand{\eq}[1]{(\ref{eq:#1})}

\newcommand{\Th}[1]{\textsc{#1}}
\newcommand{\mr}[2]{\multirow{#1}{*}{#2}}
\newcommand{\mc}[2]{\multicolumn{#1}{c}{#2}}
\newcommand{\tb}[1]{\textbf{#1}}
\newcommand{\ch}{\checkmark}

\newcommand{\red}[1]{{\color{red}{#1}}}
\newcommand{\blue}[1]{{\color{blue}{#1}}}
\newcommand{\green}[1]{{\color{green}{#1}}}
\newcommand{\gray}[1]{{\color{gray}{#1}}}

\newcommand{\citeme}[1]{\red{[XX]}}
\newcommand{\refme}[1]{\red{(XX)}}

\newcommand{\fig}[2][1]{\includegraphics[width=#1\columnwidth]{fig/#2}}
\newcommand{\figh}[2][1]{\includegraphics[height=#1\columnwidth]{fig/#2}}


\newcommand{\tran}{^\top}
\newcommand{\mtran}{^{-\top}}
\newcommand{\zcol}{\mathbf{0}}
\newcommand{\zrow}{\zcol\tran}

\newcommand{\ind}{\mathbbm{1}}
\newcommand{\expect}{\mathbb{E}}
\newcommand{\nat}{\mathbb{N}}
\newcommand{\zahl}{\mathbb{Z}}
\newcommand{\real}{\mathbb{R}}
\newcommand{\proj}{\mathbb{P}}
\newcommand{\prob}{\mathbf{Pr}}
\newcommand{\normal}{\mathcal{N}}

\newcommand{\mif}{\textrm{if}\ }
\newcommand{\other}{\textrm{otherwise}}
\newcommand{\minimize}{\textrm{minimize}\ }
\newcommand{\maximize}{\textrm{maximize}\ }
\newcommand{\st}{\textrm{subject\ to}\ }

\newcommand{\id}{\operatorname{id}}
\newcommand{\const}{\operatorname{const}}
\newcommand{\sgn}{\operatorname{sgn}}
\newcommand{\var}{\operatorname{Var}}
\newcommand{\mean}{\operatorname{mean}}
\newcommand{\trace}{\operatorname{tr}}
\newcommand{\diag}{\operatorname{diag}}
\newcommand{\vect}{\operatorname{vec}}
\newcommand{\cov}{\operatorname{cov}}
\newcommand{\sign}{\operatorname{sign}}
\newcommand{\prj}{\operatorname{proj}}

\newcommand{\softmax}{\operatorname{softmax}}
\newcommand{\clip}{\operatorname{clip}}

\newcommand{\defn}{\mathrel{:=}}
\newcommand{\peq}{\mathrel{+\!=}}
\newcommand{\meq}{\mathrel{-\!=}}

\newcommand{\floor}[1]{\left\lfloor{#1}\right\rfloor}
\newcommand{\ceil}[1]{\left\lceil{#1}\right\rceil}
\newcommand{\inner}[1]{\left\langle{#1}\right\rangle}
\newcommand{\norm}[1]{\left\|{#1}\right\|}
\newcommand{\abs}[1]{\left|{#1}\right|}
\newcommand{\frob}[1]{\norm{#1}_F}
\newcommand{\card}[1]{\left|{#1}\right|\xspace}
\newcommand{\diff}{\mathrm{d}}
\newcommand{\der}[3][]{\frac{d^{#1}#2}{d#3^{#1}}}
\newcommand{\pder}[3][]{\frac{\partial^{#1}{#2}}{\partial{#3^{#1}}}}
\newcommand{\ipder}[3][]{\partial^{#1}{#2}/\partial{#3^{#1}}}
\newcommand{\dder}[3]{\frac{\partial^2{#1}}{\partial{#2}\partial{#3}}}

\newcommand{\wb}[1]{\overline{#1}}
\newcommand{\wt}[1]{\widetilde{#1}}

\def\xssp{\hspace{1pt}}
\def\ssp{\hspace{3pt}}
\def\msp{\hspace{5pt}}
\def\lsp{\hspace{12pt}}

\newcommand{\cA}{\mathcal{A}}
\newcommand{\cB}{\mathcal{B}}
\newcommand{\cC}{\mathcal{C}}
\newcommand{\cD}{\mathcal{D}}
\newcommand{\cE}{\mathcal{E}}
\newcommand{\cF}{\mathcal{F}}
\newcommand{\cG}{\mathcal{G}}
\newcommand{\cH}{\mathcal{H}}
\newcommand{\cI}{\mathcal{I}}
\newcommand{\cJ}{\mathcal{J}}
\newcommand{\cK}{\mathcal{K}}
\newcommand{\cL}{\mathcal{L}}
\newcommand{\cM}{\mathcal{M}}
\newcommand{\cN}{\mathcal{N}}
\newcommand{\cO}{\mathcal{O}}
\newcommand{\cP}{\mathcal{P}}
\newcommand{\cQ}{\mathcal{Q}}
\newcommand{\cR}{\mathcal{R}}
\newcommand{\cS}{\mathcal{S}}
\newcommand{\cT}{\mathcal{T}}
\newcommand{\cU}{\mathcal{U}}
\newcommand{\cV}{\mathcal{V}}
\newcommand{\cW}{\mathcal{W}}
\newcommand{\cX}{\mathcal{X}}
\newcommand{\cY}{\mathcal{Y}}
\newcommand{\cZ}{\mathcal{Z}}
\newcommand{\cPi}{\mathcal{\pi}}

\newcommand{\vA}{\mathbf{A}}
\newcommand{\vB}{\mathbf{B}}
\newcommand{\vC}{\mathbf{C}}
\newcommand{\vD}{\mathbf{D}}
\newcommand{\vE}{\mathbf{E}}
\newcommand{\vF}{\mathbf{F}}
\newcommand{\vG}{\mathbf{G}}
\newcommand{\vH}{\mathbf{H}}
\newcommand{\vI}{\mathbf{I}}
\newcommand{\vJ}{\mathbf{J}}
\newcommand{\vK}{\mathbf{K}}
\newcommand{\vL}{\mathbf{L}}
\newcommand{\vM}{\mathbf{M}}
\newcommand{\vN}{\mathbf{N}}
\newcommand{\vO}{\mathbf{O}}
\newcommand{\vP}{\mathbf{P}}
\newcommand{\vQ}{\mathbf{Q}}
\newcommand{\vR}{\mathbf{R}}
\newcommand{\vS}{\mathbf{S}}
\newcommand{\vT}{\mathbf{T}}
\newcommand{\vU}{\mathbf{U}}
\newcommand{\vV}{\mathbf{V}}
\newcommand{\vW}{\mathbf{W}}
\newcommand{\vX}{\mathbf{X}}
\newcommand{\vY}{\mathbf{Y}}
\newcommand{\vZ}{\mathbf{Z}}

\newcommand{\va}{\mathbf{a}}
\newcommand{\vb}{\mathbf{b}}
\newcommand{\vc}{\mathbf{c}}
\newcommand{\vd}{\mathbf{d}}
\newcommand{\ve}{\mathbf{e}}
\newcommand{\vf}{\mathbf{f}}
\newcommand{\vg}{\mathbf{g}}
\newcommand{\vh}{\mathbf{h}}
\newcommand{\vi}{\mathbf{i}}
\newcommand{\vj}{\mathbf{j}}
\newcommand{\vk}{\mathbf{k}}
\newcommand{\vl}{\mathbf{l}}
\newcommand{\vm}{\mathbf{m}}
\newcommand{\vn}{\mathbf{n}}
\newcommand{\vo}{\mathbf{o}}
\newcommand{\vp}{\mathbf{p}}
\newcommand{\vq}{\mathbf{q}}
\newcommand{\vr}{\mathbf{r}}
\newcommand{\Vs}{\mathbf{s}}
\newcommand{\vt}{\mathbf{t}}
\newcommand{\vu}{\mathbf{u}}
\newcommand{\vv}{\mathbf{v}}
\newcommand{\uu}{\mathbf{u}}
\newcommand{\cc}{\mathbf{c}}
\newcommand{\vw}{\mathbf{w}}
\newcommand{\vx}{\mathbf{x}}
\newcommand{\vy}{\mathbf{y}}
\newcommand{\vz}{\mathbf{z}}

\newcommand{\vone}{\mathbf{1}}
\newcommand{\vzero}{\mathbf{0}}

\newcommand{\valpha}{{\boldsymbol{\alpha}}}
\newcommand{\vbeta}{{\boldsymbol{\beta}}}
\newcommand{\vgamma}{{\boldsymbol{\gamma}}}
\newcommand{\vdelta}{{\boldsymbol{\delta}}}
\newcommand{\vepsilon}{{\boldsymbol{\epsilon}}}
\newcommand{\vzeta}{{\boldsymbol{\zeta}}}
\newcommand{\veta}{{\boldsymbol{\eta}}}
\newcommand{\vtheta}{{\boldsymbol{\theta}}}
\newcommand{\viota}{{\boldsymbol{\iota}}}
\newcommand{\vkappa}{{\boldsymbol{\kappa}}}
\newcommand{\vlambda}{{\boldsymbol{\lambda}}}
\newcommand{\vmu}{{\boldsymbol{\mu}}}
\newcommand{\vnu}{{\boldsymbol{\nu}}}
\newcommand{\vxi}{{\boldsymbol{\xi}}}
\newcommand{\vomikron}{{\boldsymbol{\omikron}}}
\newcommand{\vpi}{{\boldsymbol{\pi}}}
\newcommand{\vrho}{{\boldsymbol{\rho}}}
\newcommand{\vsigma}{{\boldsymbol{\sigma}}}
\newcommand{\vtau}{{\boldsymbol{\tau}}}
\newcommand{\vupsilon}{{\boldsymbol{\upsilon}}}
\newcommand{\vphi}{{\boldsymbol{\phi}}}
\newcommand{\vchi}{{\boldsymbol{\chi}}}
\newcommand{\vpsi}{{\boldsymbol{\psi}}}
\newcommand{\vomega}{{\boldsymbol{\omega}}}

\newcommand{\rLambda}{\mathrm{\Lambda}}
\newcommand{\rSigma}{\mathrm{\Sigma}}

\newcommand{\vLambda}{\bm{\rLambda}}
\newcommand{\vSigma}{\bm{\rSigma}}

\makeatletter
\newcommand*\bdot{\mathpalette\bdot@{.7}}
\newcommand*\bdot@[2]{\mathbin{\vcenter{\hbox{\scalebox{#2}{$\m@th#1\bullet$}}}}}
\makeatother

\makeatletter
\DeclareRobustCommand\onedot{\futurelet\@let@token\@onedot}
\def\@onedot{\ifx\@let@token.\else.\null\fi\xspace}

\def\eg{\emph{e.g}\onedot} \def\Eg{\emph{E.g}\onedot}
\def\ie{\emph{i.e}\onedot} \def\Ie{\emph{I.e}\onedot}
\def\cf{\emph{cf}\onedot} \def\Cf{\emph{Cf}\onedot}
\def\etc{\emph{etc}\onedot} \def\vs{\emph{vs}\onedot}
\def\wrt{w.r.t\onedot} \def\dof{d.o.f\onedot} \def\aka{a.k.a\onedot}
\def\etal{\emph{et al}\onedot}
\makeatother

\begin{abstract}
The extraordinary electric vehicle (EV) popularization in the recent years has facilitated research studies in alleviating EV energy charging demand.Previous studies primarily focused on the optimizations over charging stations' (CS) profit and EV users' cost savings through charge/discharge scheduling events.
In this work, the random behaviours of EVs are considered, with EV users' preferences over multi-CS characteristics modelled to imitate the potential CS selection disequilibrium. 
A price scheduling strategy under decentralized collaborative framework is proposed to achieve EV shunting in a multi-CS environment, while minimizing the charging cost through multi-agent reinforcement learning. The proposed problem is formulated as a Markov Decision Process (MDP) with uncertain transition probability.
\end{abstract}

\begin{IEEEkeywords}
electric vehicle price scheduling, electric vehicle shunting, multi-agent reinforcement learning, centralised training decentralised execution, markov decision process
\end{IEEEkeywords}

\section{\uppercase{Introduction}}
\label{sec:introduction}
\begin{figure*}[ht]
  \centering
\includegraphics[width=0.62\textwidth, height = 0.30\textwidth]{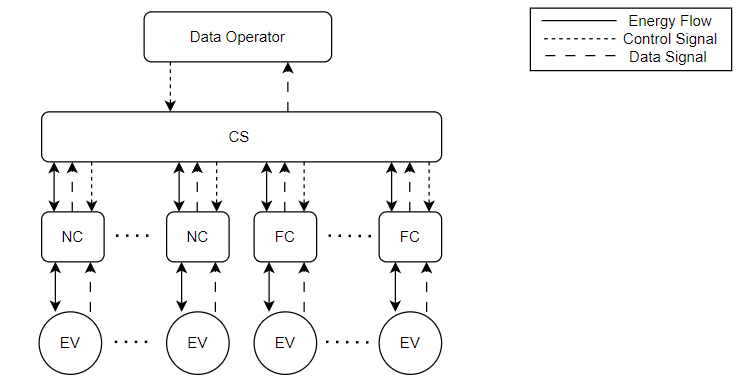}  
\caption{Decentralized collaborative framework for a single CS.}
\label{fig:single}
 \end{figure*}
According to the report from the House of Common Library ~\cite{netzero}, UK has proposed and adopted two sets of energy consumption strategies to reach the net zero policy by 2050. The vehicle carbon emissions have occupied the largest proportion of UK emission of 23\textperthousand  \ in 2022. Within the same year, UK achieved the second largest electric vehicle (EV) sales in Europe ~\cite{powerup}. The convincing results further propelled the policies over petrol vehicle prohibition and EV mandate, in turn facilitated the EV popularization thus the ever-growing charging demand. 
Correspondingly, developing solutions to alleviate the demand crisis and reduce charging peak load has gained noticeable attentions in the recent years. 
\newline
The nature of the aforementioned problems requires sequential decision-making approaches, thus implying them into Markov Decision Process (MDP) ~\cite{song2000optimal} is favoured. Reinforcement learning (RL) as one of the main stream machine learning paradigms has been a hot topic within this domain, due to its capability of learning optimal policies in  complicated environments in a model-free manner. RL is frequently applied in managing the EV charging, discharging schedulings to optimize charging costs  ~\cite{li2023constrained} ,~\cite{wen2015optimal}; In ~\cite{wen2015optimal}, RL algorithm is used to learn device based MDP models, with the implementation of a Q-table to estimate the Q function. At the other hand,  ~\cite{li2023constrained} uses recurrent deep deterministic policy gradient (RDDPG) algorithm, which proven to have great scalability in solving large-scaled MDP problems.
\newline
While the above methods focus on single agent learning problems, the real-world environment involves multi-EVs charging under single or multi-CSs. The scenarios where multi-CS are participated in, the inherent competitive and cooperative natures among CSs should also be considered. To adapt with the corresponding increased complexity, the emergence of multi-agent reinforcement learning (MARL) based approaches opens up a new realm worth investigation ~\cite{iqbal2019actor},~\cite{li2023decentralized},~\cite{zou2022intelligent},~\cite{sousa2015multi}. MARL based methods can be generalized into two categories 1) centralized execution methods ~\cite{zou2022intelligent},~\cite{sousa2015multi}, 2) decentralized execution methods ~\cite{iqbal2019actor},~\cite{li2023decentralized}.
For centralized training and decentralized execution (CTDE), ~\cite{iqbal2019actor} proposes an actor-attention-critic approach to boost reward optimization performance through an attention critic allowing dynamic agents selections at each training time point; and in ~\cite{li2023decentralized}, a decentralized collaborative framework is proposed to account situations where CS has various types of charging piles, i.e. normal charging (NC) and fast charging (FC).
\newline
Although the aforementioned methods are sufficient in providing charging cost optimization improvements, there weren't many works on modelling and, or guiding EV users' characteristics. One of the branches conducts research within the direction of EV charging navigation ~\cite{qian2019deep},~\cite{xing2022graph}, where the primary objective is to suggest users with the shortest routes between EVs and CSs, while achieving simultaneous minimization over EV users’ traveling times and charging costs. In ~\cite{qian2019deep}, a deterministic shortest charging route model (DSCRM) is utilized to extract state features from the stochastic traffic, and RL is applied to obtain an optimal strategy for the navigation problem with a practical case study in Xi’an’s real-sized zone.  Whereas in ~\cite{xing2022graph}, the features are extracted through graph convolutional transformation and learnt through graph reinforcement learning (GRL), the optimization goal is achieved with the aid of a modified rainbow algorithm.
Other directions consider modelling the EV users' preferences~\cite{wang2015ev},~\cite{liu2022pricing}.~\cite{wang2015ev} implements a user preference and price based charging algorithm, experimented in the UCLA parking lots; and ~\cite{liu2022pricing} proposed a modified probability model. 
\newline
In this work, the decentralized collaborative framework is extended to include multi-CSs, the CS characteristics are represented in an asymmetric manner, i.e. diversified charging prices at different time periods,  varied charging pile sizes. To study the EV users' behaviours, a probabilistic EV user preference model is built accounting real time charging prices, EV to CS distances and CS sizes (pile numbers) . Hence, a normalized linear distance model is also developed along with the random arrivals of the EVs. Finally, since the probabilistic EV user preferences can potentially result in an asymmetric selection over a certain CS, a price scheduling strategy is proposed to allow CSs to compete over prices and attract EV users in real time through shared CS information, and eventually achieve EV shunting, relieving the stress of the potential space congestion and charging demands over certain individual CS. 

The main contributions of this paper are summarized as follows:
\begin{enumerate}

\item A multi-CS based decentralized collaborative framework is proposed, where each pile with an EV attached is treated as an agent, which is able to execute local and independent charging, discharging actions; the charging problem is formulated as MDP. A Value-Decomposition Network (VDN) ~\cite{sunehag2017value} MARL algorithm is utilized to perform charging cost optimization. 

\item A probabilistic EV user preference model is designed, which considers real time prices, EV-CS distances and CS sizes to predict EV users selections over various CSs; each term is assigned with a tuned weight, together with a linear distance model developed to normalize and quantify the distance parameter.

\item A real time dynamic price scheduling strategy is proposed, by utilizing the shared information between the stations, individual CS is able to influence EV users' selections through live price competition; the reduced price for the influenced EV is reclaimed in the remaining time steps before the departure to ensure CS profit, while achieving EV shunting.

\item Through VDN and Q-mix ~\cite{rashid2020monotonic}  performance evaluations, the results show that our price scheduling strategy is sufficient in producing substantial average reward improvements and faster episode convergence over the baseline.
\end{enumerate}

The rest of the paper is structured as follows. \autoref{sec:related} presents the concept of single CS based decentralized collaborative framework. \autoref{sec:method} proposes the multi-CS based collaborative framework, followed with the modified probabilistic EV user preference model definition, and the MDP problem formulation. \autoref{sec:implementation} explains the proposed price scheduling strategy procedure and the training phase. In \autoref{sec:experiments}, rigorous evaluations were carried out to validate the effectiveness of the proposed method, together with the corresponding optimisation performances. Finally, \autoref{sec:conclusion} concludes our work. 
\section{\uppercase{Background}}
\label{sec:related}

\subsection{Decentralized collaborative framework (single CS)}
\begin{figure*}[ht]
  \centering
\includegraphics[width=0.80\textwidth, height = 0.32\textwidth]{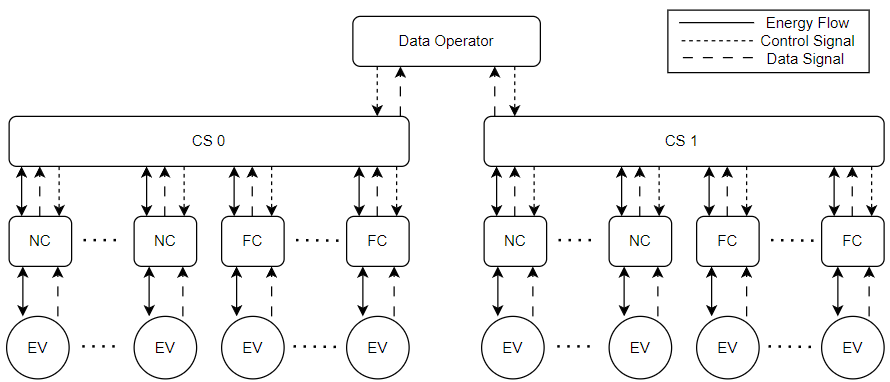}  
\caption{ Multi(dual)-CS based decentralized collaborative framework  }
  \label{fig:multics}
 \end{figure*}

Under a single CS based decentralized collaborative framework, the CS contains multiple charging piles, including both NC and FC, as shown in \autoref{fig:single}. Once EV arrives and attaches itself to an allocated pile, its information, i.e. current state of charge (SOC), time of arrival $t_{a}$, estimate time of departure $t_{d}$ and the charging demand $e_{d}$ will be transmitted via the pile and CS to the data operator; presented as data signals.Since each pile has the ability to make local and independent charging, discharging decisions (as forms of energy flow exchanges), data signals also embed these decisions together with the CS real time charging prices. In the single CS scenario, the data operator's task is to gather these data and send control signals to ensure the accumulated independent charging decisions will not cause CS capacity overload , this is implemented through live monitoring of the total CS power demand at each time step, in the same manner as the procedure proposed in ~\cite{li2023constrained}. In our work, the single CS based framework is extended to a dual-CS decentralized collaborative framework, as the result, the data operator also acts as a medium between the CSs to exchange information of the charging prices, real time charged EV number of each CS, allowing the proceeding of the proposed price "competition" algorithm; the detailed descriptions are presented in \autoref{sec:implementation}.

\section{\uppercase{environment formulation}}
\label{sec:method}
\subsection{Multi(dual)-CS based collaborative framework}
From the single-CS based  framework mentioned in \autoref{sec:related}, we propose a multi-CS based collaborative framework by extending the single station into dual stations, as depicted in \autoref{fig:multics}. Under the new framework, CS 0 and CS 1 are differentiated by their 1) Charging pile number $N_{z}$ , with $z\in[0..1]$. 2) Real time charging price  $p_{t,z}$. The data operator now carries additional responsibility of integrating and exchanging the pile number and price information between the CSs. 
The proposed framework inherits the same concepts as its baseline, in which each arrived EV, equivalent as a pile under operation, is treated as an agent, moreover, the optimization objective is to minimize the accumulated charging cost $C_{acc}$, through each individual pile's charging decisions, expressed as 
\begin{equation}
    C_{acc} = \sum_{z} \sum_{t} p_{t,z}
\end{equation}

\subsection{Probabilistic EV user preference model}
Inspired by the works from ~\cite{zhao2020deployment} and ~\cite{liu2022pricing}, we propose a modified probabilistic model to imitate EV users' selection preferences over the CSs. The following parameters being used are adopted from ~\cite{zhao2020deployment}.
\newline
The selection probability $Pr_{i,z,t_a^{i}}$ of the $i^{th}$ arriving EV selecting CS $z$, at its arrival time $t_a^{i}$ can be formally defined as follows: 
\begin{equation}
    Pr_{i,z,t_a^{i}}  = \frac{U_{i,t_a^{i},z}}{\sum_{z=0}^1 U_{i,t_a^{i},z}}
\end{equation}
where $U_{i,t_a^{i},z}$ is the EV user utility function, in turn is directly proportional to the attractiveness $Att_{i,t_a^{i},z}$ of CS $z$ on the $i^{th}$ EV user . The user utility function can be expressed as 
\begin{equation}
    U_{i,t_a^{i},z}  = \frac{Att_{i,t_a^{i},z}}{T_{i,z}}
\end{equation}
where $T_{i,z}$ defines the closeness between EV $i$ and CS $z$, represented as 
\begin{equation}
    T_{i,z}  = t_{c} + t_{i,z}
\end{equation}
with $t_{c}$ being the average EV charging time and $t_{i,z}$ being the travelling time from EV $i$ to CS $z$.
\newline
In a practical scenario, we consider four major factors that have significant influences on EV users' decisions: 1) Charging pile number $N_{z}$. By considering EV users' risk stop loss characteristics, the arrived users' selections would lean towards the CS with higher pile number, due to the likelihood of having more available spaces, and the smaller chances of discovering fully occupied station at arrival, which results in longer distance travelled, towards inferior charging locations. 2) Arrival time charging price $p_{t_a^{i},z}$. EV users generally would prefer cheaper prices, hence, the CS that offers cheaper prices will gain more attractions from the users. 3) EV $i$ and CS $z$ travel distance $d_{i,z}$. Longer travel distance will make CS less attractive to the EV users. 4) EV arrival state of charge $SOC_{a}$. High $SOC_{a}$ will increase CS $z$'s attractiveness. Furthermore, we define a SOC threshold $SOC_{th}$, when $SOC_{a}$ is less than $SOC_{th}$, the EV is urgent to get charged, leading to less attraction under large $d_{i,z}$, and vice versa. This term will be discussed in more details in the upcoming session. 
\newline
Based on the aforementioned factors, the attractiveness $Att_{i,t_a^{i},z}$ can be formulated as a linear weighing function: 
\begin{equation}
    Att_{i,t_a^{i},z}  = w_{0}\cdot N_{z} -w_{1}\cdot p_{z,t_a^{i}} - w_{2}\cdot d_{i,z} + w_{3}\cdot (SOC_{a}-SOC_{th})
\label{eq:att}
\end{equation}
$w_{j}$ with $j\in[0..3]$ are the weighing coefficients, the negative sign indicates degrading effect on the user attraction. 

\begin{figure}[h]
  \centering
\includegraphics[width=0.45\textwidth, height = 0.08\textwidth]{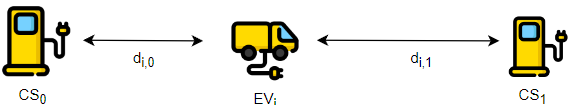}  
\caption{Linear EV-CS distance model}
  \label{fig:EVMODEL}
 \end{figure}

\subsection{Distance normalization model}
To illustrate the concept of utilizing distance normalization, we first propose a linear EV-CS distance model, as shown in  \autoref{fig:EVMODEL}. $d_{i,0}$ and $d_{i,1}$ are the corresponding distances between EV $i$ and CS 0, 1, and they are assumed to be the shortest routes by default.
Recalling from equation \eq{att}, high $SOC_{a}$ would result in higher station attraction to EV users, and the SOC is related to the EV-CS distances. But due to the unit differences (SOC has values ranging from 0 to 1 while the actual distance $d_{i,z}$ is significantly larger in comparison),  they are separated and assigned with different weighing coefficients $w_{2}$ and $w_{3}$, in turn increases the attraction function complexity and obscures the relationship between them.
\newline
Hence, we divide the original $d_{i,z}$ by the total distance from EV $i$ to both stations, and obtain the normalized distance $\widetilde{d_{i,z}}$, expressed as 
\begin{equation}
	\widetilde{d_{i,z}} = \frac{d_{i,z}}{d_{i,0}+d_{i,1}}  , z\in [0..1]
\end{equation}
As a result, the original attractiveness $Att_{i,t_a^{i},z}$ can be modified into $\widetilde{Att}_{i,t_a^{i},z}$, as shown the following formula:
\begin{equation}
	\widetilde{Att}_{i,t_a^{i},z} =  w_{0}\cdot N_{z} -w_{1}\cdot p_{t_a^{i},z} - w_{2}\cdot (\widetilde{d_{i,z}}+SOC_{th}-SOC_{a})
\label{eq:newatt}
\end{equation} 
From equation \eq{newatt}, $w_3$ is eliminated as  $\widetilde{d_{i,z}}$,  $SOC_{a}$ and  $SOC_{th}$ now share a highly overlapped value range. When the arrival SOC exceeds the threshold, the observation indicates the EV having sufficient energy to travel relative long distances, in turn alleviates the attraction degradation caused from $\widetilde{d_{i,z}}$, and vice versa. 
\newline
Additionally, the traveling time $t_{i,z}$ can be calculated using $\widetilde{d_{i,z}}$:
\begin{equation}
	 t_{i,z} = \frac{\widetilde{d_{i,z}}} {\widetilde{v_{i}}}
\end{equation}
where $\widetilde{v_{i}}$ is the normalized vehicle velocity of EV $i$.
\subsection{Markov decision process}
As a discrete-time stochastic control process, MDP is sufficient in modeling our charging,discharging decision problems, due to the sequential feature of the decision making process. MDP consists of four components: 1) State. It describes the agent's (arrived EV, or the pile under operation)  current situation, and the future 'next' state only depends on the present state ,owing to the MDP's definition. 2) Action. The charging, discharging decisions made by the NC, FC piles, under the current state time step. 3) State transition function. The function that evaluates the probability in which the  action under current state causing the future 'next' state transition. 4) Reward function. The  optimization goal when the action causes current state transition to the future 'next' state, and it's directly related to the accumulated charging cost.
\newline
In this section, we use MDP to model our decision problem, with detailed definitions presented in the following subsections. 

\subsubsection{State}
The state at time step $t$ can be defined as $s_{i,t,z} = (s_{i,t,z}^{pile},s_{t,z}^{station})$, where $s_{i,t, z}^{pile}$ is the state of pile $i$ in CS $z$ at time $t$, which can be formally defined as
\begin{equation}
	s_{i,t,z}^{pile} = (SOC_{i,t,z}, P^{max}_{i,t,z},P^{min}_{i,t,z},t^{stay}_{i,t,z})
\end{equation} 
where $t^{stay}_{i,t,z}$ is the remaining staying time of the $i^{th}$ EV in CS $z$ at time step $t$, defined as 
\begin{equation}
	t^{stay}_{i,t,z} = t^{d}_{i,t,z} - t^{cur}
\end{equation} 
with $t^{d}_{i,t,z}$ being the time of departure,  and $t^{cur}$being the current time step.  
\newline
Furthermore, $s_{t,z}^{station}$ is the state of CS $z$ at time $t$, and can be expressed as 
\begin{equation}
	s_{t,z}^{station} = (t^{cur}, p_{t,z},eme^{total}_{t,z})
\end{equation} 
where $eme^{total}_{t,z}$ is CS $z$'s total emergency at time $t$, can also in turn be represented as 
\begin{equation}
   eme^{total}_{t,z} =   \sum_{i}^{N_{i,t,z}} eme_{i,t,z} 
\end{equation} 
with $N_{i,t,z}$ being the number of EVs attached to CS $z$ at time $t$, and $eme_{i,t,z}$ being the charge emergence of the individual EV $i$ at CS $z$ and time step $t$, as shown in \eq{eme}.
\begin{equation}
     eme_{i,t,z}  =
    \begin{cases}
      \frac {(SOC_{i,t,z}^{need}-SOC_{i,t,z})\cdot C_{i,z}}{t^{stay}_{i,t,z}} & \text{if $SOC_{i,t,z}$  $<$  $SOC_{i,t,z}^{need}$}\\
      0  & \text{otherwise}
      
    \end{cases}  
\label{eq:eme}
\end{equation} 
$SOC_{i,t,z}^{need}$ is the SOC needed for EV $i$ at time $t$ charging at station $z$ before its departure. 
\subsubsection{Action}
In this section we define the $i^{th}$ pile's action in CS $z$ at time step $t$ as $a_{i,t,z}$, within a range of $[-1,1]$. $a_{i,t,z}$ represents the corresponding charging,discharging decision made by the pile.
Thus, the real time charging power $P^{real}_{i,t,z}$ is expressed as 
\begin{equation}
   P^{real}_{i,t,z} = \frac {(a_{i,t,z} + 1)}{2} * (P^{max}_{i,t,z} - P^{min}_{i,t,z}) + P^{min}_{i,t,z}
\end{equation} 
where $P^{max}_{i,t,z}$ and $P^{min}_{i,t,z}$ are the corresponding maximum and minimum pile charging,discharging powers at CS $z$ at time $t$.As a result, we defined the total charging power $P^{total}_{t,z}$ in CS $z$, time step $t$ as  
\begin{equation}
   P^{total}_{t,z} = \sum_{i} P^{real}_{i,t,z}
\end{equation}

\subsubsection{State transition}
 To model the probability of state transition from $s_{t}$ to $s_{t+1}$ under the influence of $a_{t}$, we formulate the state transition as 
\begin{equation}
   s_{t+1} =  f(s_{t},a_{t},\omega_{t})
\end{equation} 
$f(\cdot)$ is the state transition function, and the term $\omega_{t}$ is defined to indicate the uncertainty of the state transition.

\subsubsection{Reward function}
The station reward $\vr^{station}_{t,z}$ for CS $z$ at time step $t$ can be formally defined as 
\begin{equation}
   \vr^{station}_{t,z} =  P^{total}_{t,z} \cdot p_{t,z}
\end{equation} 
To ensure the total charging power $P^{total}_{t,z}$ does not exceed the maximum power limit $P^{max}_{z}$ for each CS $z$, we define a punishment term $\vr^{punish}_{t,z}$ as
\begin{equation}
     \vr^{punish}_{t,z}  =
    \begin{cases}
      P^{max}_{z} - P^{total}_{t,z} & \text{if $P^{max}_{z}$ $<$  $P^{total}_{t,z}$}\\
      0 & \text{otherwise}
    \end{cases}  
\end{equation} 
Therefore, the total reward at each time step $t$ is expressed as 
\begin{equation}
   \vr_{t} = \sum_{z=0}^1 \vr^{c}_{t,z} + A \cdot \vr^{punish}_{t,z}
\end{equation} 
Where $A$ is a scaling factor.

\begin{algorithm}[!h]

 \caption{EV price scheduling training procedure}
 \label{alg:train}

 \SetKwInOut{Input}{input}
 \SetKwInOut{Output}{output}
 \Input{state $s_{i,t,z} = (s_{i,t,z}^{pile},s_{t,z}^{station})$}
 

\Output{VDN network parameters}
 {
 Randomly initialize VDN network parameter $\theta$.\\
 Copy VDN network parameter to target network $\theta'=\theta$.\\
 \For{episode $m = 1:M$ }{\For{$t = t_{start}:t_{end}$}{\For{CS $z$ = CS 0 : CS 1}{\For{pile $i$ = pile 0 : pile $n$}{obtain EV $i$'s state $s_{t,z} = (s_{i,t,z}^{pile},s_{t,z}^{station})$\\
 \If{$p_{i-1,t^{cur}-1,0} \neq 0$, $p_{i-1,t_{cur}-1,1} \neq 0$}
 {$p_{i-1,t-1,0} + \alpha $, $p_{i-1,t-1,1} + \beta $.
   }
\If{$N_{t^{cur},0}$ $<$  $N_{t^{cur},1}$}
{ $p_{i,t^{cur}+1,z}  = p_{t^{cur},0} - \alpha$}
\If{$N_{t^{cur},0}$ $>$  $N_{t^{cur},1}$}
{ $p_{i,t^{cur}+1,z}  = p_{t^{cur},0} - \beta$}
select action $a_{i,t,z}$ based on $\epsilon-$greedy policy

 }
 calculate reward $\vr^{station}_{t,z} =  P^{total}_{t,z} \cdot p_{t,z}$\\
 sample transition $<s_{i,t,z},a_{i,t,z},r_{t,z},s_{i,t+1,z}>$ in $B_{1}$\\
 }calculate total reward $\vr_{t} = \sum_{z=0}^1 \vr^{c}_{t,z} + A \cdot \vr^{punish}_{t,z}$}}
sample mini batch $<s_{i,z},a_{z},r_{z},s_{i+1,z}>$ from $B_{1}$\\
set $ y^{tot}=  r_{i,z} + \gamma \cdot max_{\va'}Q_{tot}(\vtau',\va',s'_{i,z};\theta')$\\
update VDN network by minimizing loss $ L_{\theta} =  \sum_{i}\sum_{z}(y_{i,z}^{tot} -  Q_{tot}(\vtau,\va,s_{i,z};\theta))^2$
}\\
\end{algorithm}

\section{\uppercase{proposed price scheduling strategy}}
\label{sec:implementation}
\subsection{Price scheduling}
\begin{figure*}[ht]
  \centering
\includegraphics[width=0.70\textwidth, height = 0.60\textwidth]{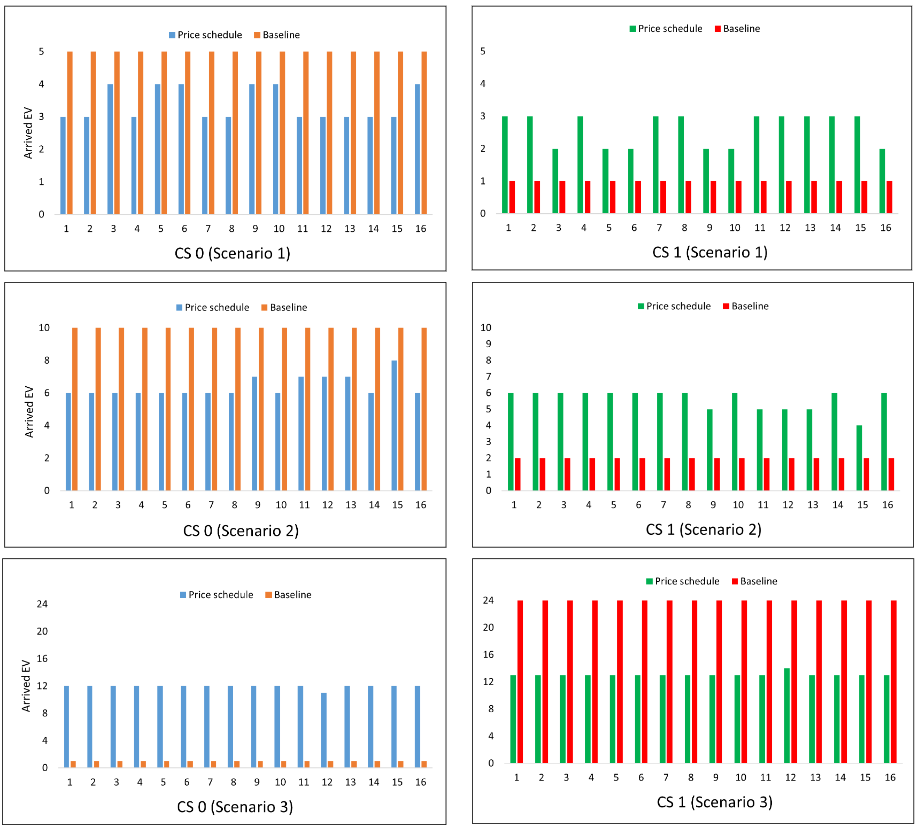}  
\caption{EV distributions over price scheduling and baseline methods, under each scenario}
\label{fig:combine}
 \end{figure*}

In this section, we propose the novel price scheduling strategy to achieve EV shunting through multi-CS price "competition" at each simultaneous time step. At $t^{cur}$, the data operator receives the pile occupation information $N_{t^{cur},0}$ and $N_{t^{cur},1}$ from each CS, if the pile occupation is imbalanced,$N_{t^{cur},0} \neq N_{t^{cur},1}$; the data operator will command the CS with lower attached EV to reduce the charging price by a sufficient amount to attract EV at step $t^{cur} + 1$, namely $p_{i,t^{cur}+1,z}$ , as shown in equation \ref{eq:price}.
\begin{equation}
     p_{i,t^{cur}+1,z}  =
    \begin{cases}
      p_{t^{cur},0} - \alpha_{t^{cur}}  & \text{if $N_{t^{cur},0}$ $<$  $N_{t^{cur},1}$}\\
      p_{t^{cur},1} - \beta_{t^{cur}}  & \text{if $N_{t^{cur},0}$ $>$  $N_{t^{cur},1}$}\\
      p_{t^{cur},z}    & \text{otherwise}
      
    \end{cases}  
\label{eq:price}
\end{equation} 
The terms $\alpha_{t^{cur}}$ and $\beta_{t^{cur}}$ are the corresponding price reductions at each CS and the current time $t^{cur}$. 
\newline
Moreover, $p_{i,t,z}$ is a subset of CS charging price $p_{t,z}$, it is only implied to those attracted EVs at time step $t$, and will be charged back at the next step $t+1$, in turn can be expressed as
\begin{equation}
     p_{i,t+1,z}  =
    \begin{cases}
      p_{t-1,0} + \alpha_{t-1}  & \text{if $z = 0$, $p_{i-N_{t-1,0}^{att},t-1,0} \neq 0$}\\
      p_{t-1,1} + \beta_{t-1}  & \text{if $z = 1$,$p_{i-N_{t-1,1}^{att},t-1,1} \neq 0$}
      
    \end{cases}  
\end{equation}
where $N_{t-1,0}^{att}$ and $N_{t-1,1}^{att}$ are the numbers of price attracted EVs to CS 0 and 1 at time step $t-1$.

\subsection{Training phase}
In our work, we use VDN ~\cite{sunehag2017value} MARL algorithm,a Q-learning based approach to resolve the proposed multi-agent reinforcement learning problem. The primary task of the network is to learn the total action value function $Q_{tot}(\vtau,\va)$; where $\vtau = (\tau_{1,0},...,\tau_{n,1})$ is observation history, with $\tau_{i,z}$ being the joint action observation history of agent $i$ from CS $z$ and $\va = (a_{1,0},...,a_{n,1})$ is the action set. $Q_{tot}(\vtau,\va)$ can be represented as a summation of each value function $Q_{i,z}(\tau_{i,z},a_{i,t,z})$: 
\begin{equation}
     Q^{tot}(\vtau,\va)= \sum_{i} Q_{i,z}(\tau_{i,z},a_{i,z};\theta_{i})  
\end{equation}
where $\theta_{i}$ is the network parameter.
\newline
The loss function of the VDN network is defined as
\begin{equation}
     L_{\theta} =  \sum_{i}\sum_{z}(y_{i,z}^{tot} -  Q_{tot}(\vtau,\va,s_{i,z};\theta))^2
\end{equation}
$y^{tot}$ is the total target action value, which is calculated as shown in equation \eq{target}.
\begin{equation}
     y^{tot}=  r_{z} + \gamma \cdot max_{\va'}Q_{tot}(\vtau',\va',s'_{i,z};\theta')
\label{eq:target}
\end{equation}
The term $\gamma$ is the discount factor, $\vtau'$, $\va'$ and $s'_{i,z}$ are the corresponding observation history, action and state after state $s_{i,z}$ takes action $\va$ and receive reward $r_{i,z}$; $\theta'$ is the parameter of the target network. 
\newline
Furthermore, we apply the $\epsilon-$greedy action selection to balance the exploration-exploitation trade off, the optimal action that maximizes the action value function is set with a probability of 1-$\epsilon$, as shown in equation \eq{epl}.
\begin{equation}
     Pr(a_{i,t,z} = a_{i,t,z}^{opt}) = 1 - \epsilon, 0 < \epsilon < 1
\label{eq:epl}
\end{equation}
Where $a_{i,t,z}^{opt}$ is the optimal action, and a random action is chosen with a probability of $\epsilon$, the transition  $<s_{i,t,z},a_{i,t,z},r_{t,z},s_{i,t+1,z}>$ at time step $t$ is stored in a replay buffer $B_{1}$. With the above information, the following price scheduling training procedure is summarised in algorithm \ref{alg:train}.

\section{\uppercase{simulations}}
\label{sec:experiments}

\subsection{simulation Setup}
\begin{figure*}[ht]
  \centering
\includegraphics[width=0.75\textwidth, height = 0.62\textwidth]{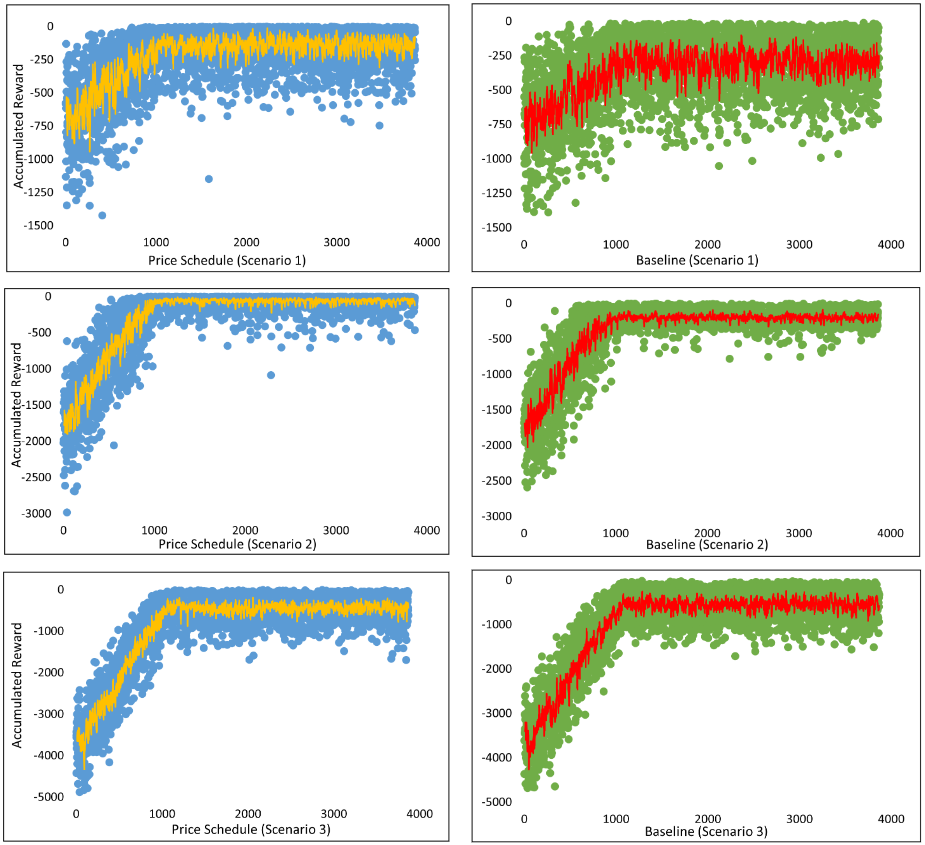}  
\caption{VDN accumulation reward distribution}
\label{fig:VDNcombine}
 \end{figure*}

\begin{figure*}[ht]
  \centering
\includegraphics[width=0.75\textwidth, height = 0.62\textwidth]{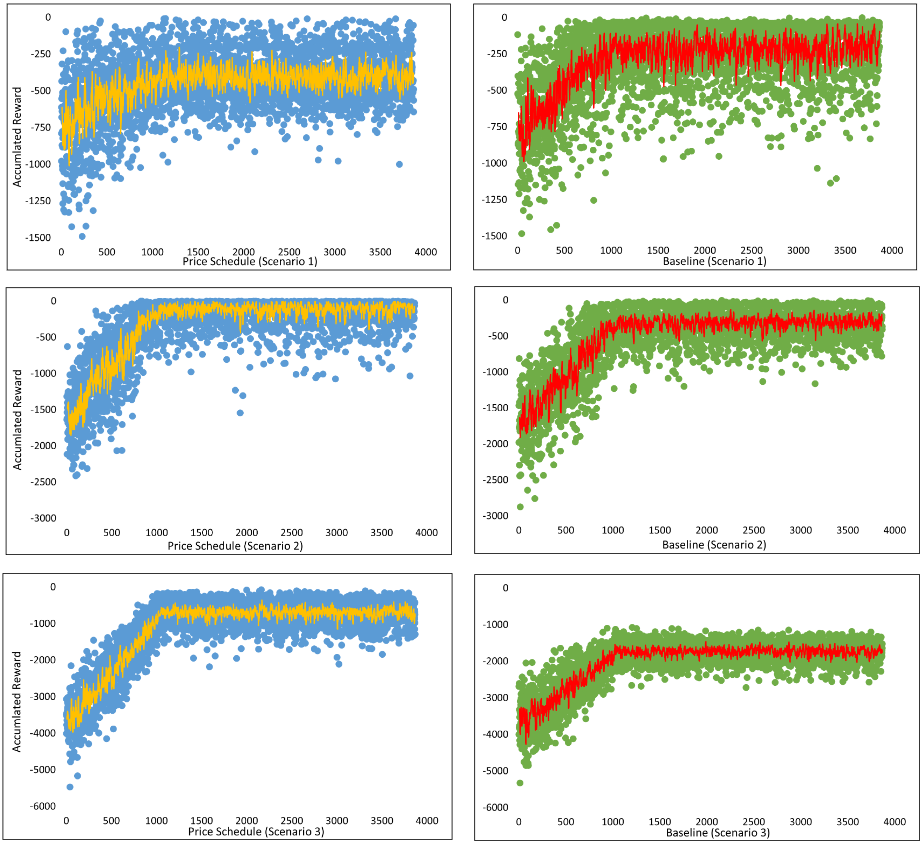}  
\caption{Q-mix accumulation reward distribution }
\label{fig:QMIXcombine}
 \end{figure*}

In our simulation set up, we adopt the Chinese time-of-use tariff for our CS 0 from \cite{li2023decentralized}, \cite{li2023constrained}. Furthermore, we developed the CS 1 tariff based on the original version, with higher prices at each corresponding time period; the tariffs for both stations are presented in \autoref{tab:soa-balanced}. 
The arrival time $t_{a}$ ,departure time $t_{d}$, arrival state of charge $SOC_{a}$ and the normalized EV-CS 0 distance $\widetilde{d_{i,0}}$ (and $\widetilde{d_{i,1}}$ = 1 - $\widetilde{d_{i,0}}$ ) all follow a normal distribution , with a corresponding boundary limit, as shown in \autoref{tab:two}. Moreover, the needed state of charge $SOC^{need}$ is set to 0.8, minimum state of charge $SOC^{min}$ and maximum state of charge $SOC^{max}$ are set to 0.2, 0.9 respectively. The maximum charging power for NC pile is 6 kWh,with a maximum battery capacity of $C^{NC}$ = 24 kWh, and for FC pile the maximum charging power is 30 kWh, with a battery capacity $C^{FC}$ = 180 kWh; the random seed of 50 is selected. 
\newline
For our VDN network model, the layers in the network have 64 neurons, and the learning rate of the network is set to be 0.001. The batch size is 32, and the buffer size is 2000.

\subsection{EV shunting evaluation}

Firstly, we evaluate the effectiveness of the price scheduling strategy, i.e. it achieves sufficient EV shunting under each case study. In this sub section, three different scenarios are generated for the shunting evaluation, where each scenario has varied numbers of EVs requiring charging and charging pile sizes at each station: 1) 6 arriving EVs, CS 0 with 3 FC piles and 2 NC piles, CS 1 with 4 FC piles and 3 NC piles; 2) 12 arriving EVs, CS 0 with 6 FC piles and 4 NC piles, CS 1 with 8 FC piles and 6 NC piles; 3) 25 arriving EVs, CS 0 with 14 FC piles and 6 NC piles, CS 1 with 20 FC piles and 10 NC piles. The weight coefficients for the attraction model under each scenario are configured as 
\begin{equation}
	  w_{0} , w_{1} , w_{2} = (0.25,3.8,0.7)
\label{eq:newatt}
\end{equation} 
Each scenario is simulated for 1000 times, and 16 samples are intercepted for representation, as shown in \autoref{fig:combine}. Under scenario 1, our price scheduling method results in two majority arrived EV size pairs of $(CS 0, CS1) = (3,3), (4,2)$, whilst without price scheduling the common size pair is $(5,1)$.Under scenario 2, the price scheduling results in common size pairs of $(6,6), (7,5)$ in comparison with the baseline of $(10,2)$; and finally for scenario 3, the result shows a difference of $(12,13)$ vs $(1,24)$.
The above results are sufficient in proving our price scheduling strategy in achieving reasonable EV shunting.
\newline

\subsection{VDN performance evaluation}
To evaluate the performance of our price scheduling strategy, we pick the VDN model performance with the identical EV user preference model (without price scheduling) as our baseline, and compare the corresponding accumulated rewards under the aforementioned scenarios, depicted in \autoref{fig:VDNcombine}.
In \autoref{tab:three}, the average reward at convergence and the episode of convergence are presented under each scenario, and results of the price scheduling method, baseline are compared. Scenario 1: Our price scheduling method converges 68 episode faster than the baseline, while the average reward at convergence of our method is 157.537 higher than the baseline, and the accumulated rewards appear to be more compact. Scenario 2: The price scheduling method converges 82 episodes faster than the baseline, and the corresponding average reward is higher by 137.77; again, the accumulated rewards are virtually more compact in comparison with the baseline results. Scenario 3: Our method converges 16 episodes faster, while achieving a significant higher average award of -442.093, with a better overall reward compactness over the baseline. 

\subsection{Q-mix performance evaluation}
To further investigate the effectiveness of the proposed method, we replace the VDN model with Q-mix network. The CS $z$ passes the agent observation $o_{i,t,z}$ and previous action step $a_{i,t-1,z}$ to the agent network, where the agent network is a recurrent neural network (RNN) that consists 3 layors: multilayer perceptron (MLP), gated recurrent unit (GRU) and  MLP. The agent network outputs each value function $Q_{i,z}(\tau_{i,z},a_{i,t,z})$ into the mixing network, where the mixing network consists of hypernetworks that take in the state $s_{t,z}$, all the generated $Q_{i,z}(\tau_{i,z},a_{i,t,z})$ together with the weights and biases from the hypernetworks are used to produce the total action value function $Q_{tot}(\vtau,\va)$.
\newline
Additionally,the local $argmax$ applied on each individual action value function should obtain the same monotonicity effect as the global $argmax$ applied on the total action value function, in which can be expressed as 
\begin{align} 
argmax_{\vu}Q_{tot}(\vtau,\va) &=                       
\begin{pmatrix}
 argmax_{u_{1,0}}Q_{1,0}(\tau_{1,0},a_{1,0})\\
\vdots  \\
argmax_{u_{n,1}}Q_{n,1}(\tau_{n,1},a_{n,1})
\end{pmatrix}
\end{align}
Moreover, this expression can be further generalized to a constraint, as shown in equation \eq{constraint}.
\begin{equation}
\frac{\partial Q_{tot}}{\partial Q_{i,z}} \geq 0, \forall i \in [1..n] , \forall z \in [0..1]
\label{eq:constraint}
\end{equation}
The Q-mix model adapts the same parameter settings as VDN. 
\newline
The same three scenarios are implied, with the corresponding accumulated reward performances shown in \autoref{fig:QMIXcombine}. 
From \autoref{tab:four}, under scenario 1 the price scheduling strategy converges 1 epoch faster over the baseline, however, the corresponding average reward at convergence is 182.975 less than the baseline and the accumulated rewards are shown to be less compact. Moving onto scenario 2, the price scheduling method converges 95 episodes slower than the baseline, but the average reward is 200.864 higher and the data scattering appears to be more compact. For scenario 3, the price scheduling method achieves a noticeably higher average award of -696.103 over the baseline, similar to the simulation results for the VDN network, while achieving it with faster episode convergence and a comparable overall reward compactness between the two methods.
 The above results indicate that, under scenarios where charging pile sizes at each station and the arrived EV sizes are sufficiently large, the price scheduling strategy is capable to producing noticeable reward optimization improvement over the baseline.
Together with the VDN performance results, it is sufficient to claim that under reasonably large dataset, the proposed price scheduling strategy is capable of providing noticeable reward optimization improvements, while achieving it with faster episode convergence.

\begin{table}
\small
\centering
\caption{\textsc{Chinese Time-of-Tariff}}

\setlength\tabcolsep{9pt}
\begin{tabular}{lccc}
\toprule
               Time   &       CS 0             &  CS 1   &      \\ \midrule
                
                                       \mc{4}{{Price(CNY/kWh)}}  \\ \midrule


07:00 - 10:00                          &  1.0044    & 1.2044    \\
11:00 - 13:00                   & 0.6950 & 0.7950\\ 

14:00 - 17:00                         &  1.0044     & 1.2044    \\
18:00 - 19:00                & 0.6950 & 0.7950\\
20:00 - 06:00                   & 0.3946 & 0.4946\\
\bottomrule
\end{tabular}
\vspace{0pt}

\label{tab:soa-balanced}
\end{table}

\begin{table}
\small
\centering
\caption{\textsc{Commuting Behaviour Distributions}}

\setlength\tabcolsep{5pt}
\begin{tabular}{lccc}
\toprule
             Parameters     &         Distribution           & Boundary  &       \\ \midrule

$t_{a} $                       & $\mathcal{N}(9,1^{2})$ &  $ 7 \leq t_{a} \leq 11 $         \\
$t_{d} $               & $\mathcal{N}(19,1^{2})$   & $17 \leq t_{d} \leq 21$ \\ 
$SOC_{a} $                        &$\mathcal{N}(0.4, 0.1^{2})$ &  
 $ 0.2 \leq SOC_{a} \leq 0.6$          \\

$\widetilde{d_{i,0}}$               & $\mathcal{N}(0.5, 0.3^{2})$   & $ 0 < \widetilde{d_{i,0}} < 1 $  \\ 

\bottomrule
\end{tabular}
\vspace{0pt}

\label{tab:two}
\end{table}

\begin{table}
\small
\centering
\caption{\textsc{VDN Performance Results}}

\setlength\tabcolsep{2pt}
\begin{tabular}{lccc}
\toprule
             Method     &         Average reward            & Episode of convergence  &       \\  \midrule

                                       \mc{4}{\Th{Scenario 1}}  \\ \midrule

Price Schedule                         &  \tb{-141.383}     & \tb{1170}    \\ 
Baseline                  & -298.92 & 1238\\ \midrule
\mc{4}{\Th{Scenario 2}}  \\ \midrule
Price Schedule                        &  \tb{-63.288}     & \tb{1063}    \\ 
Baseline                 & -201.058 & 1145 \\ \midrule
\mc{4}{\Th{Scenario 3}}  \\ \midrule
Price Schedule                         &  \tb{-442.093}     & \tb{1176}     \\
Baseline                & -547.313  &  1192\\

\bottomrule
\end{tabular}
\vspace{0pt}

\label{tab:three}
\end{table}

\begin{table}
\small
\centering
\caption{\textsc{Q-MIX Performance Results}}

\setlength\tabcolsep{2pt}
\begin{tabular}{lccc}
\toprule
              Method    &       Average reward             & Episode of convergence &       \\ \midrule

                                       \mc{4}{\Th{Scenario 1}}  \\ \midrule

Price Schedule                         &  -401.474     & \tb{1214}   \\ 
Baseline                  & \tb{-218.499} & 1215 \\ \midrule
\mc{4}{\Th{Scenario 2}}  \\ \midrule
Price Schedule                        &  \tb{-112.736}     & 1128     \\ 
Baseline                 & -313.6 & \tb{1223}\\ \midrule
\mc{4}{\Th{Scenario 3}}  \\ \midrule
Price Schedule                         &  \tb{-696.103}     & \tb{1096}     \\
Baseline                & -1732.4 & 1142 \\

\bottomrule
\end{tabular}
\vspace{0pt}

\label{tab:four}
\end{table}

\section{\uppercase{conclusion}}
\label{sec:conclusion}

To conclude, we introduced a multi-CS decentralized collaborative framework, while employing a modified EV user preference model. Furthermore, we proposed a price scheduling method that achieves comparable EV shunting performance. With numerous simulations over two separate CSs and under three different scenarios , the proposed method is proven to achieve sufficient reward optimisation and convergence improvement results over the baseline. For potential future research, we aim to introduce multi-neural network interface, with the current price scheduling strategy reinforced with additional networks to improve the capability of handling large scaled EV transportation models. Moreover, the optimization goal can be extended, such as merging CS route recommendation goals with price reduction, adapting further practicalities in a real world situation.

\bibliographystyle{ieeetr}
{\small
\bibliography{reference}}

\end{document}